\documentclass[aps,prl,twocolumn,groupedaddress]{revtex4-2}
\usepackage{aas_macros}
\usepackage{amsmath}
\usepackage{tikz}
\usepackage{bm}
\usepackage{makecell}

\usepackage{xcolor}
\usepackage[version=4]{mhchem}
\usepackage{comment}

\begin{document}


\title{Self-mediation of runaway electrons via self-excited wave-wave and wave-particle interactions}

\author{Qile Zhang}
\email{qlzhanggo@gmail.com}
\affiliation{University of Maryland, College Park, Maryland, 20742, USA}
\affiliation{Los Alamos National Laboratory, Los Alamos, NM 87545, USA}

\author{Yanzeng Zhang}
\affiliation{School of Nuclear Science and Technology, University of Science and Technology of China, Hefei, Anhui 230027, China}
\affiliation{Los Alamos National Laboratory, Los Alamos, NM 87545, USA}

\author{Qi Tang}
\affiliation{School of Computational Science and Engineering, Georgia Institute of Technology, Atlanta, GA 30332, USA
}
\affiliation{Los Alamos National Laboratory, Los Alamos, NM 87545, USA}

\author{Xian-Zhu Tang}
\email{xtang@lanl.gov}
\affiliation{Los Alamos National Laboratory, Los Alamos, NM 87545, USA}

\begin{abstract}
Nonlinear dynamics of runaway electron induced wave instabilities can
significantly modify the runaway distribution critical to tokamak
operations. Here we present the first-ever fully kinetic simulations
of runaway-driven instabilities towards nonlinear saturation in a warm
plasma \textcolor{black}{where collisional damping is subdominant}. It
is found that the slow-X modes grow an order of magnitude faster than
the whistler modes, and they parametrically decay to produce whistlers
much faster than those directly driven by runaways. These
parent-daughter waves, as well as secondary and tertiary wave
instabilities, initiate a chain of wave-particle resonances that
strongly diffuse runaways to the backward direction.  This reduces
almost half of the current carried by high-energy runaways, over a
time scale orders of magnitude faster than experimental shot
duration. These results beyond quasilinear analysis may impact
anisotropic energetic electrons broadly in laboratory, space and
astrophysics.

\end{abstract}


\maketitle

\textbf{Introduction.---}
One of the most efficient ways to generate relativistic
electrons in a dilute plasma is runaway acceleration by a strong
electric field along the magnetic
field,~\cite{Dreicer59,connor-hastie-NF75} coupled with an avalanche
growth mechanism due to knock-on collisions between primary runaways
and background cold
electrons~\cite{Sokolov:1979,Jayakumar:1993,Rosenbluth97}. Plasma wave
instabilities excited by these relativistic runaway
electrons~\cite{Parail-NF-1978,Pokol2014PhPl,Aleynikov2015} and
their roles in modifying the runaway electron distribution through
nonlinear wave-particle interaction, have piqued long-standing
interest from both a basic plasma physics perspective and the
practical need of mitigating runaway electrons in tokamak plasmas. The
latter comes about because the runaways can cause severe damage on the
plasma-facing components during both tokamak
startup~\cite{deVries-nf-2019,deVries-ppcf-2020,Hoppe-etal-JPP-2022}
and major
disruptions~\cite{Hender:2007,Boozer:2015,breizman2019physics}, which
presents a critical challenge for tokamak power
reactors~\cite{eidietis-fst-2021,creely-etal-PoP-2023}.
Outside magnetic fusion, runaway electrons can also form during solar flares \cite{Alaoui2021apj}.
Interaction of energetic electrons and their self-induced waves plays
critical roles in regulating the transport and heat flux induced by
these energetic electrons, for example, in Earth's
magnetosphere \cite{Yu-etal-FASS-2023}, solar flares
\cite{Roberg-Clark2019ApJ} and astrophysical intracluster medium
\cite{Roberg-Clark2018prl}. To facilitate these and similar
applications in a variety of laboratory, space, and astrophysical
plasmas, we must understand the basic plasma physics of runaway-wave
interaction and its nonlinear saturation.

Recent experimental advances in diagnosing the runaway electron
distribution, via, for example, spatial, temporal, and energetically
resolved measurement of bremsstrahlung hard-x-ray emission, provide
information on the energy and pitch dependence of the runaway electron
distribution~\cite{Paz-Soldan2017prl}. Direct measurement of
high-frequency electromagnetic waves in tokamak experiments supplied
the evidence of runaway-induced plasma wave
instabilities~\cite{Spong2018prl}. These hardware advances offer an
unprecedented opportunity to contrast predictions from theory and
simulations with experimental
observations~\cite{Paz-Soldan2017prl,Spong2018prl,Liu2018prl}. The
most remarkable success to date has been on the role of
forward-propagating (with respect to the runaway direction) whistler
waves that are excited by runaways via the anomalous Doppler-shifted
cyclotron
resonances~\cite{Aleynikov2015NucFu,Liu2018prl,Breizman2023a_pop}.
This finding can be contrasted with the physical picture that
extraordinary waves above the whistler branch, also known as the
slow-X modes \cite{Ram2000pop}, can be excited by the runaway
electrons via the same resonance~\cite{Pokol2014PhPl, Komar2012JPhCS}.
Most intriguingly, these authors~\cite{Pokol2014PhPl, Komar2012JPhCS}
also found that being of much higher frequency than the whistler
branch, the slow-X modes could have much higher growth rates and
stronger quasilinear pitch angle diffusion, from an analysis using a
model runaway distribution.  \textcolor{black}{More recent
  analysis~\cite{Zhang2025pop_excitation} reveals
  that even in a collisional plasma, the slow-X modes remain the far
  more robust instability compared with the whistlers, despite their
  higher collisional damping rates.~\cite{Aleynikov2015NucFu}} The
instrumentation limitation in previous DIII-D
experiments~\cite{Spong2018prl} prevents direct measurement of the
primary whistler modes, let alone the even higher frequency
X-modes. This leaves these two distinct physical scenarios unresolved:
one dominated by whistler instability and the other by slow-X modes.

Further complicating the situation, the saturation physics of these
runaway wave instabilities were previously examined using the quasilinear
theory, for both whistler
and slow-X branches~\cite{Liu2018prl,Pokol2014PhPl}. Common concerns for quasilinear
saturation analysis include (1) the mischaracterization of saturated
states if nonlinear coupling is the dominant mechanism; (2) even for
systems that saturate at the marginal stability boundary, inclusion of
parametric decay instability and secondary/tertiary instability
associated with an evolving distribution function, often neglected or
incomplete, can be essential for accuracy but it is not known \textit{a
priori}; and (3) the quasilinear diffusion approximation can be
problematic.  First-principles nonlinear kinetic simulation is thus a
necessary examination for physics fidelity and may guide the
improvement of quasilinear analysis if it applies at all.

Here, for the first time, fully kinetic particle-in-cell simulations
are successfully deployed to study runaway self-driven instabilities
toward nonlinear saturation, initiated by a self-consistent runaway
distribution from a drift-kinetic solver. We find that the slow-X
modes grow an order of magnitude faster than the whistler modes,
confirming an intriguing feature previously noted in
Ref.~\cite{Pokol2014PhPl, Komar2012JPhCS}, and they go through
parametric decay to produce whistlers much faster than those directly
driven by runaways. More interestingly, the slow-X waves can initiate
a chain of wave-particle resonances that strongly diffuse runaways to
the opposite (backward) direction at moderate and high energy, which
occurs much faster than the time scales of collisional current
dissipation and runaway acceleration.  These backward diffusion
processes strongly modify the runaway distribution and reduce almost
half of the runaway current. The new physics findings significantly
modify what is known in the literature on runaway-wave dynamics
mentioned above.

\textbf{Numerical methods.---} We deploy the typical tokamak start-up
parameters $J_{RE}=2MA/m^2, n_e=0.6\times10^{19}m^{-3}, T_e=320eV, B=1.45T$
with $|\omega_{ce}|/\omega_{pe}=1.84$. From a relativistic drift-kinetic
Fokker-Planck-Boltzmann (FPB)
solver~\cite{guo2017phase,guo-mcdevitt-tang-pop-2019}, we compute the
runaway electron distribution in the runaway avalanche regime with a
strong electric field $E=65E_c$ (with $E_c$ the Connor-Hastie field
\cite{Connor1975NF}). Such an electric field is quite reasonable in a
start up scenario \cite{deVries2023NucFu} \textcolor{black}{or during an adequately mitigated ITER disruption\cite{Zhang2025pop_excitation}. Here we do not consider those mitigation scenarios that aim for a low $E/E_c,$ where runaways are either avoided or minimized.~\cite{McDevitt-Tang-NF-2023}}  The momentum space
distribution of runaways has a low energy boundary at $p=3m_e v_{te}$
that matches onto a bulk Maxwellian-J\"untter distribution with
$v_{te}=\sqrt{2T_e/m_e}$ the electron thermal speed of the background
plasma. When the runaway avalanche exponentially increases the runaway
current to the total current, the resulting runaway distribution is
fed into the fully kinetic VPIC code \cite{Bowers2008} to study the
self-induced instabilities and wave-particle interactions on a much
faster time scale compared to the small- and large-angle collisions,
and radiation damping.

\textcolor{black}{By using a rather large value of $J_{RE}=2MA/m^2,$ the
  simulation setup mimics an interesting runaway-mitigation scenario in which the
  background plasma was sufficiently cold that the runaway-driven
  modes were collisionally stabilized at full runaway current, but a
  rapid background plasma heating (via, for example, Electron Cyclotron Heating or other
  external sources) suddenly lowers the collisional damping rate so
  the runaway-driven wave instabilities are liberated for self-mediation of runaways. For a normal
  runaway current ramp-up, the simulation corresponds to an idealized
  setup that takes a snap shot of the nonlinear dynamics by which
  runaway electron distribution saturates into near-marginality by
  self-excited wave instabilities on time scale much faster than the
  runaway current ramp-up.  In this case, the large $J_{RE}$ value is favored
  to enable feasible computational costs, for the higher growth
  rates of the modes at larger $J_{RE},$ with the understanding that a
  lower $J_{RE}$ in the earlier phase of runaway current ramp-up has the
  same or similar slow-X  modes~\cite{Zhang2025pop_excitation}.  }

The VPIC simulations use proton-electron plasma with the realistic
proton-electron mass ratio. The temperature $T_e=320eV$ corresponds to
a thermal-to-light speed ratio $v_{te}/c=0.035$. The grid size is
$\Delta_x=0.0125d_e = 0.5 \lambda_{de}$, with $d_e=c/\omega_{pe}$ the
electron inertial length and $\omega_{pe}$ the electron plasma
frequency, and $\lambda_{de}$ the electron Debye length. The time step
is $dt\omega_{pe}=0.01$. Considering the huge difference in particle
number densities between the thermal and runaway electrons  ($n_{re}=0.0082n_e$), we employ the
weighted macro-particle approach for the thermal (with $p<3m_e
v_{te}$) and runaway (with $p>3m_e v_{te}$) electrons. Specifically,
we represent the runaway tail population with a 10 times smaller
macro-particle weight compared to the thermal electrons so that the
macro-particle number for runaways is enhanced by 10 times for better
statistics. 2700 macro-particles per cell are used for the thermal
electrons. As a simplified setup, the PIC simulation includes one
spatial dimension with periodic boundary, three velocity
dimensions, and an initially uniform magnetic field and plasma. This
corresponds to the tokamak magnetic axis without the effect of
trapped electrons. Since the distribution carries a parallel current
$J_{RE}=2MA/m^2$, to be consistent with the uniform field, we Lorentz boost
all electrons opposite to the runaway direction by the averaged
parallel velocity $v_d=0.007351c$ to cancel the current. Since $v_d\ll
v_{te}$, the effect of this boost on the electron distribution is
minimal. To make wave modes sufficiently continuous over $k$ as in
reality, we use a long enough periodic domain size $L_x=1344d_e$ to
ensure a small wave mode spacing $\Delta k=2 \pi/L_x$. The spatial
dimension is at an angle $\theta$ to the magnetic fields,
which is chosen as $\theta=40^{\circ}$. Our linear dispersion
analysis following Ref.~\cite{Aleynikov2015NucFu,Liu2018prl} shows that, for such
$\theta=40^\circ$, the growth rate of slow-X modes
($\sim4\times10^{-3}\omega_{pe}$) \textcolor{black}{can reach about the  maximum} and is an order of magnitude larger than
that of the fastest whistler mode ($\sim10^{-4}\omega_{pe}$), which is
excited at $\theta\sim70^{\circ}$. We run the simulation till the
distribution saturates, which is at $t\omega_{pe}\sim 10^6$. The
collisional damping time scales \cite{Aleynikov2015NucFu} of the
relevant waves ($t\omega_{pe}\sim 10^7-10^8$) are much longer than the whole simulation, so we can
neglect collisions.

\begin{figure*}
\includegraphics[width=\textwidth]{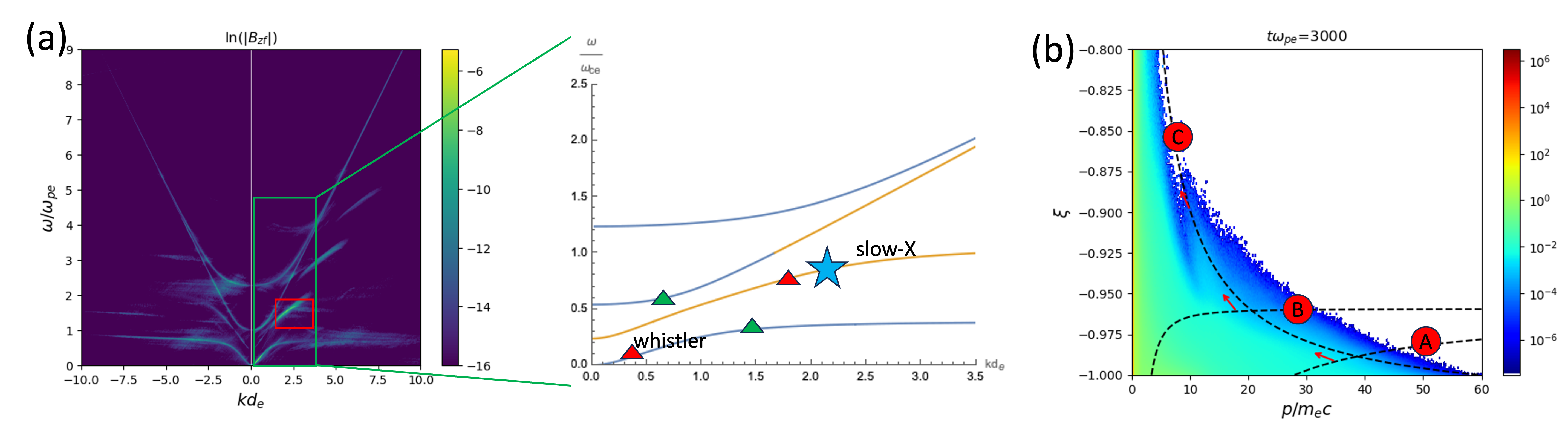}
\caption{ (a): the Fourier space of magnetic field $B_z$ where slow-X
  waves are strongly driven (red box). A schematic picture of a
  zoom-in window (green box) on the Fourier space from cold plasma
  dispersion~\cite{Stix1992wavesbook} illustrates different branches
  (especially the whistler and slow-X), and different waves on the
  branches. The strongly driven parent slow-X mode (blue star) can
  parametrically decay into two pairs of daughter wave groups (red
  or green triangles), both including whistler waves. (b): in the momentum
  space distribution, the slow-X waves diffuse the high-energy tail
  over pitch and momentum at this early time, as shown by the
  resonance lines (dashed lines) and diffusion directions (red
  arrows).\label{slowx_fourier_decay}}
\end{figure*}

\textbf{Strong slow-X mode drive and its parametric decay at short time scale.---} \label{sec:slow-X}
Based on the dispersion analysis, the high-energy runaway tail
($p/m_ec\sim30$) can drive waves through $n=1$ anomalous Doppler
resonance not only on the whistler branch \cite{Liu2018prl, Fülöp2006pop} but also
on the slow-X branch \cite{Komar2012JPhCS,Pokol2014PhPl}. Here the
resonance condition reads
\begin{equation}\label{res_condition}
    \omega-k_\parallel V \xi=n \omega_{ce}/\gamma,
\end{equation}
where $\omega_{ce}<0$ for the negative charge. For $\theta=40^{\circ}$, the maximum growth
rate for the slow-X branch is $\sim4\times 10^{-3}\omega_{pe}$, much
higher than the whistler branch $\sim10^{-5}\omega_{pe}$. As shown in
the Fourier space at very early time $t\omega_{pe}=5000$ in Fig.
\ref{slowx_fourier_decay}(a), the amplitude of slow-X waves in the red
box has grown large, with even nonlinearly generated higher harmonics
that have multiples of $\omega$ and $k$~\cite{shivamoggi2012nonlinear_waves_book}.

Once the strongest forward propagating slow-X mode grows to a large
amplitude, it can go through parametric decay to produce two pairs of
lower frequency modes (e.g., see Fig.~\ref{slowx_fourier_decay}(a)),
including forward whistler waves.
Specifically, the strongest slow-X wave (blue star,
$\omega=0.86|\omega_{ce}|=1.58\omega_{pe},kd_e=2.16$) parametrically decays into two pairs
of wave groups (red or green triangles). The red triangles include a
low frequency whistler wave
($\omega\sim0.07|\omega_{ce}|=0.13\omega_{pe},kd_e\sim0.35$), and a high frequency
slow-X wave ($\omega\sim0.8|\omega_{ce}|=1.47\omega_{pe},kd_e\sim1.88$) near the parent
wave. The green triangles can also produce
whistler waves at higher $\omega$ and $k$ ($kd_e\sim1-2$). These parametric decay processes are much faster than the whistler
modes directly driven by the runaways, i.e. the primary whistler modes. These daughter whistlers observe high amplitude and broad spectrum. 


We will explore the runaway dynamics in the momentum space over pitch $\xi$ and momentum $p$. The diffusion direction of runaway electrons in the local momentum space by an individual resonant wave (satisfying Eq.~(\ref{res_condition})) can originate from the directional
gradient of the runaway distribution $f(p,\xi)$ as $\hat{L}f$, where
\begin{align}\label{L_operator}
    \hat{L}&=\frac{1}{p}\frac{\partial}{\partial p}-\frac{1}{p^2}\frac{n\omega_{ce}/\gamma-\omega(1-\xi^2)}{\omega\xi}\frac{\partial}{\partial \xi}\\ \nonumber
    &=-(-\frac{1}{p},\frac{1}{p^2}(\xi-k_\parallel v/\omega))\cdot(\frac{\partial}{\partial p},\frac{\partial}{\partial \xi}).
\end{align}
Since the wave is driven by the gradient $\hat{L}f$ and the quasilinear diffusion of $f$ is given by $\hat{L}^2f$  \cite{Liu2018prl,Stix1992wavesbook}, the diffusion direction can be defined by the unit vector 
\begin{equation}\label{eq:diffusion_direction}
    \hat{\mathbf{g}}=(-\frac{1}{p},\frac{1}{p^2}(\xi-k_\parallel v/\omega))/\|(-\frac{1}{p},\frac{1}{p^2}(\xi-k_\parallel v/\omega))\|*sgn(\hat{L}f)
\end{equation} 
This vector in fact represents the particle flux direction when the
wave is smoothing out the gradient $\hat{L}f$. It must be noted that the runaway
electrons can either lose (diffused towards small $p$, positive $\hat{L}f$) energy to or gain (towards large $p$, negative
$\hat{L}f$) from the resonant wave. The former corresponds to runaways
driving the wave. The latter will cause the wave damping by
the runaways, in which case the wave must be driven by other mechanisms
(e.g., the parametric decay) or by different resonances. We will
use two arrow colors (red and blue) to denote the wave
gaining/losing energy from the wave-particle interaction in the local momentum space.

\begin{figure*}
\includegraphics[width=.8\textwidth]{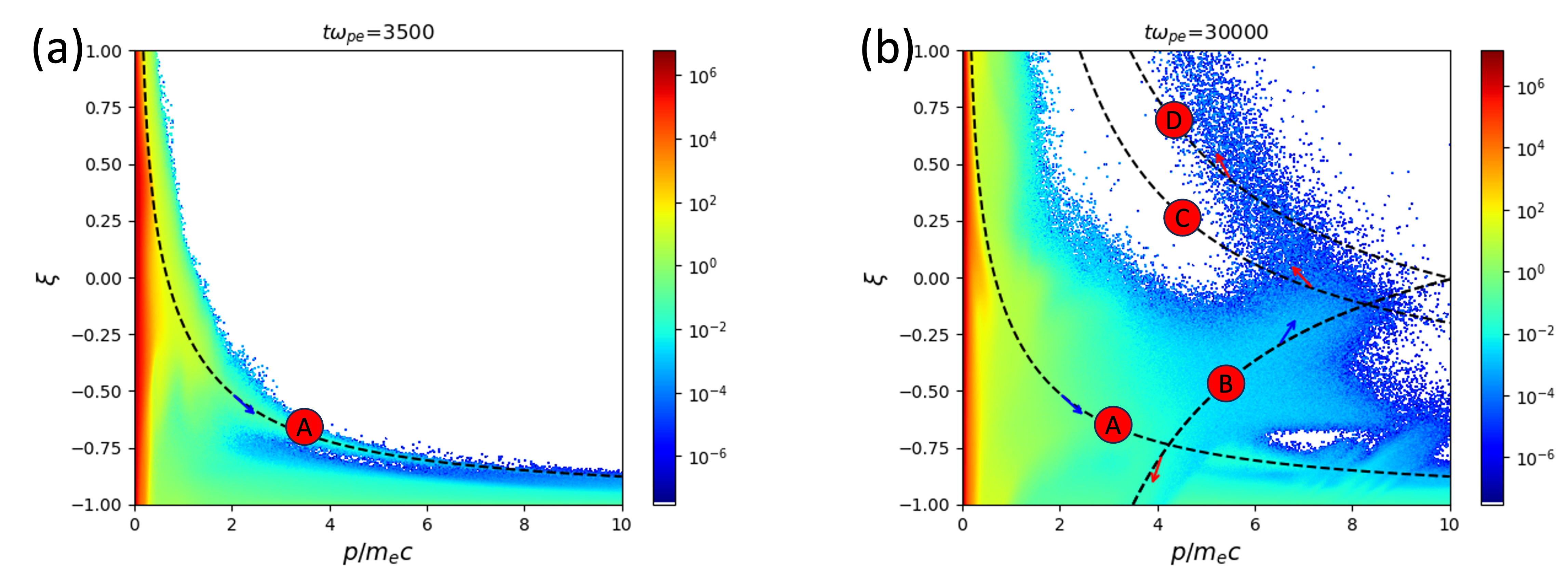}
\caption{The momentum space at moderate energy at different times. The
  strong slow-x waves initiate a chain of wave-particle resonances
  labeled as ABCD that diffuse runaways to the backward direction
  along their diffusion directions.
\label{backward_diffuse_ABCD}}
\end{figure*}

During the short time scale $t\omega_{pe}<5000$, the slow-X modes,
including the parent and daughter waves, can notably diffuse the high
energy runaway tail ($p>10m_ec$), which initiates the fast
backward diffusion.  Fig. \ref{slowx_fourier_decay}(b) shows zoom-in high energy
runaway tail distribution close to $\xi=-1$. The high-energy electrons are being diffused subsequently along diffusion directions
on different resonance lines. Specifically, from resonance lines A to
C, they involve resonances $n=1$ and $n=0$ of the parent slow-X wave
($\omega=1.58\omega_{pe}, kd_e=2.16$), and $n=-1$ of the daughter
slow-X wave ($\omega=1.47\omega_{pe}, kd_e=1.88$). They lead to a
finger in $f(p,\xi)$ towards lower energy and higher pitch, which
drives primary and secondary slow-x modes.  Meanwhile, these
strong parent and daughter slow-X modes also accelerate electrons,
from the edge of the backward thermal bulk, along the resonance line
of $n=-1$ to higher energy as shown in
Fig.~\ref{backward_diffuse_ABCD}(a) (label A for mode
$\omega=1.53\omega_{pe}, kd_e=2$). This establishes a strong finger in
$f(p,\xi)$, by damping all the slow-X waves. This
extended finger will contribute to the forward current at moderate
energy.

\textbf{Fast backward diffusion at moderate energy at medium time scale.---}\label{sec:moderate_energy_diffuse}
During the medium time scale ($t\omega_{pe}\sim3\times 10^4$), the strong
finger from the damping of slow-X modes can initiate a chain of
wave-particle interactions through $n=-1$ resonance, which can diffuse
moderate energy runaways ($p/m_ec\sim5$) to the backward direction
(e.g., see Fig. \ref{backward_diffuse_ABCD}(b)).  The strong finger
first provides free energy to trigger a series of secondary backward
propagating whistler waves (visible in the Fourier space in
Fig. \ref{slowx_fourier_decay}(a)) through $n=-1$ resonance (e.g., the
red arrow at resonance line B for $\omega=0.2\omega_{pe}, kd_e=
0.47$). These secondary whistlers can further diffuse the runaways
towards higher pitch (the blue arrow on B), forming another
finger. This new finger sequentially encounters the resonance lines of
$n=-1$ of the forward whistler waves from the parametric decay of
slow-X such as: label C ($\omega=0.28\omega_{pe}, kd_e= 0.61$) and
label D ($\omega=0.18\omega_{pe}, kd_e= 0.45$), which diffuse runaway
electrons straight to the backward direction.

\begin{figure*}
\includegraphics[width=\textwidth]{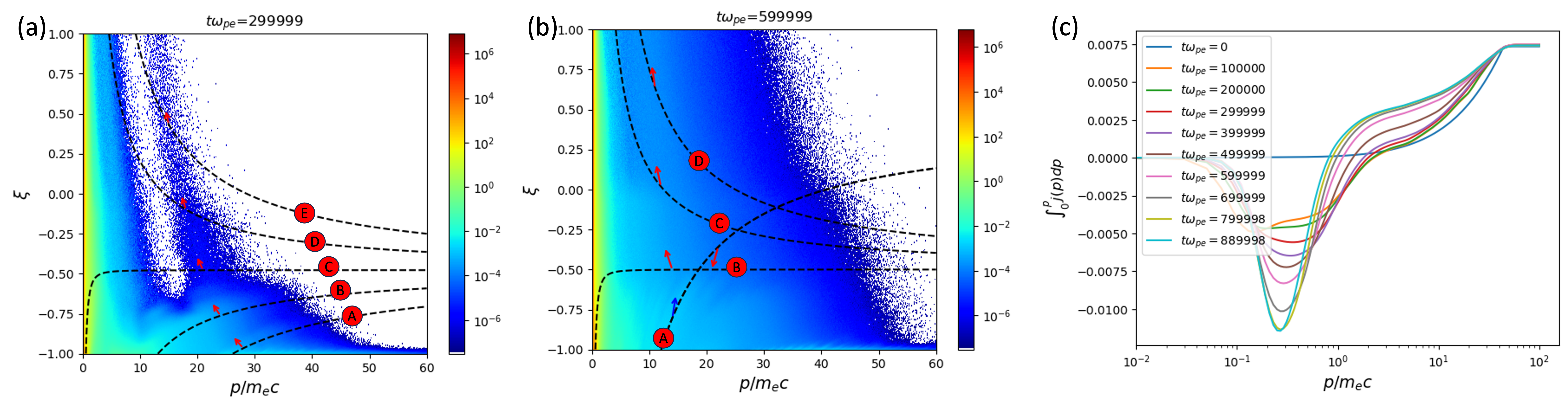}
\caption{ (a): in the high-energy momentum space, the backward
  diffusion occurs sequentially along the diffusion directions of
  multiple resonances of forward whistler waves. (b) at later time the
  triggered backward whistlers diffuse electrons to higher pitch to
  encounter resonances of forward whistlers. Eventually the backward
  diffusion significantly fills the high pitch momentum space at high
  energy. (c): the integrated current distribution over momentum, where
  almost half of the integrated current at high energy is converted to
  lower energy during this
  process. \label{backward_diffuse_highenergy}}
\end{figure*}

\textbf{Fast backward diffusion at high energy at long time scale.---}\label{sec:high_energy_diffusion}
Following the fast backward diffusion of high-energy runaways by the slow-X modes, the whistler waves produced from the parametric decay process of the slow-X mode can continue to backward diffuse the high-energy runaway
tail through a chain of resonances over a long time scale
$t\omega_{pe}\sim3\times 10^5$, as shown in Fig.~\ref{backward_diffuse_highenergy}. Specifically,
Fig. \ref{backward_diffuse_highenergy}(a) shows the backward diffusion
of runaways on different resonance lines of an example forward whistler wave
$\omega=0.13\omega_{pe}, kd_e=0.35$ near the peak of the whistler
spectrum from the parametric decay, sequentially with $n=2$ to $-2$
resonances from label A to E. This forms a strong diffusion finger
straight to backward. The quasi-linear diffusion
\cite{GuoZ2018pop,Stix1992wavesbook} of these different harmonic
resonances can be connected to each other by the broad spectrum of
whistler waves from parametric decay (see diffusion coefficients in
the supplemental material). The backward diffusion to lower energy
also allows the excitation of secondary whistler waves,
which further enhance the diffusion at later time.

The distinct backward finger formed at high energy introduces free
energy to trigger a series of tertiary backward
whistler waves through $n=-1,-2$ resonances.
An example resonance line of $n=-2$ in
Fig. \ref{backward_diffuse_highenergy}(b) (label A) is shown for a
backward whistler mode $\omega=0.09\omega_{pe}, kd_e=0.28$, {which is
  driven by the high-energy runaways at $p/m_ec\sim22$ and
  $\xi\sim-0.35$ (the red arrow)}. Once it is
excited, it will diffuse electrons of $p/m_ec\sim14$ (the blue arrow)
to higher pitch to encounter the broad spectra of resonance lines of
$n=0,-1,-2$ (label B to D) of the forward whistler waves
(e.g. $\omega=0.15\omega_{pe}, kd_e=0.38$). Eventually, the high-pitch
momentum space at high energy ($p/m_ec\ge10$) is significantly filled.
See also the supplemental movie
demonstrating all the fast backward diffusion
processes. Fig.~\ref{backward_diffuse_highenergy}(c) shows the
evolution of the integrated current distribution over momentum during
this process.  We have reversed the previous Lorentz boost of $v_d$ to
retrieve the current before the integration. The current contained at
the high-energy runaway tail decreases significantly over time as the
average pitch of high-energy electrons increases.  When the current
profile eventually saturates, almost half of the high energy current
(e.g. $p/m_ec>10$, above 5MeV) is converted to be carried by lower
energy superthermal electrons at $p/m_ec\lesssim1$. This whole strong
process occurs at a fast time scale of $3\times 10^5
\omega_{pe}^{-1}\sim10^{-6}s$, which is extremely short compared to
experimental shot duration or the collisional runaway current damping
time scale $\tau_c=4\pi\epsilon_0^2 m_e^2 c^3/e^4 n_e
ln\Lambda\sim0.37s$, with $ln\Lambda$ the Coulomb Logarithm.
Note that the superthermal current at $0.3<p/m_ec\lesssim1$ significantly increases over time, contributed by multiple processes. While the early time superthermal current is significantly contributed by the strong finger extending to moderate energy from the $n=-1$ damping of slow-X waves, at the later time it is contributed by both $n=0$ landau damping of the forward whistlers and $ n=-1$ damping of the backward whistlers. Interestingly, this increasing superthermal current results in parallel electric fields that push the thermal bulk backward due to current conservation (Ampere's law), leading to negative integrated current from the bulk electrons with $p/m_ec<0.2$. 

\textbf{Discussion.---}
First fully kinetic simulations of the excitation and nonlinear saturation
of runaway-electron-driven electromagnetic wave instabilities reveal
a qualitatively new physics picture of complex wave-wave interactions
and runaway-wave-interaction-induced secondary and tertiary wave
instabilities, in contrast to previous quasilinear
analysis emphasizing only the primary instability. The slow-X modes,
found to be the fastest growing instabilities, can
parametrically drive both high and low frequency whistlers, often
dominating over the primary whistler modes that are directly driven by
the runaways. Wave-particle resonant interaction develops rapidly
evolving features in momentum space for the electron distribution
function, including particle acceleration, slowing down, and strong pitch diffusion. The
secondary and tertiary wave instabilities are sequentially excited to
facilitate rapid pitch spread to backward for runaways and the growth of
superthermal electrons, with the net result of quickly transferring substantial plasma
current from high-energy runaways to medium-energy runaways and
superthermal electrons, which is a form of runaway mitigation that limits the runaway energy and increases its dissipation.


As notable experimental signatures, \textcolor{black}{the most obvious
  is high-frequency signal in the slow-X mode range. Frequencies
  higher than $|\omega_{ce}|$ are already observed by
  \cite{Almagri2023APSabstract}. One can also expect both forward and
  backward electromagnetic waves (\textcolor{black}{also observed by
    \cite{Almagri2023APSabstract}})} at large amplitudes, particularly
the whistler branch, as well as strong chirping. Similarly, a large
population of relativistic electrons can be measured to move in the
opposite direction of the original distribution, i.e. in the
co-current direction. \textcolor{black}{While backward relativistic
  electrons have not been explicitly investigated in experiments,
}\textcolor{black}{wall deposition of both forward and backward
  subcritical energetic electrons has been observed in runaway
  experiments\cite{Beidler2024wall}. \textcolor{black}{From our
    studies, these subcritical energetic electrons with
    $\gamma\lesssim 1,$ moving in the backward direction, could be
    populated by self-excited waves as shown in Figure
    \ref{backward_diffuse_ABCD}}}. Although the current simulations do
not account for magnetic trapping, the same wave-runaway interaction
physics should lead to a significant trapped high-energy electron
population through wave-induced pitch angle scattering. This can
provide a robust drive for Alfv\'en waves in the MHz range that were
\textcolor{black}{already} observed in experiments and thought to be
driven by processional drift resonance with trapped runaways
\cite{Lvovskiy2018ppcf,Lvovskiy2019nf,Liu2023prl}.

While we initialize the kinetic simulations with the
collisional-time-scale FPB solution to explore these fast-time-scale
wave dynamics \textcolor{black}{that saturate the runaway electron
  distribution toward near-marginality}, how the waves and
distributions self-consistently couple for long-time runaway current
evolution remains to be further explored, \textcolor{black}{for the
  prohibitively high computational cost with a fully kinetic
  approach.}  \textcolor{black}{Although the current simulation deploys
  a large $J_{RE}$ for faster growing instabilities and
  quicker nonlinear saturation, we note that the physics insights
  should translate to lower $J_{RE}$ cases in the earlier stage of
  runaway current ramp-up, when the initial robust instability growth
  relaxes the runaway distribution toward near-marginality.  This is
  because the same slow-X modes remain the most unstable ones at lower
  $J_{RE}$~\cite{Zhang2025pop_excitation}. The
  same matching/resonance conditions for parametric decay and
  wave-particle interaction would constrain the saturation dynamics
  along a similar path as described.  Similarly, we have
  reduced the complexity of the simulation by restricting it to a
  single angle of $\theta\sim40^\circ,$ which allows the strongest
  wave-wave interaction between the primary slow-X and the
  low-frequency whistler waves (which can resonate with high-energy
  electrons). For different angles, the daughter waves involved in
  parametric decay instabilities will vary.  \textcolor{black}{The uncovered physics of primary-wave excitation, parametric decay, daughter-wave resonances, etc. will similarly apply to a 2D3V system. In reality, waves would also bounce radially to reach outer regions of lower temperature with higher collisional damping, but the slow-X mode is shown to be the primary instability even in a colder temperature \cite{Zhang2025pop_excitation}. } The multi-angle physics
  and} the physical effects of tokamak geometry such as trapped
electrons, spatial dependence and radial transport need to be explored
with simulations of higher dimensions.  The revealed basic processes
of fast backward diffusion facilitated by the slow-X waves may
strongly impact not only the runaway electron dynamics in tokamaks,
but also likely the anisotropic energetic electron evolution and
transport broadly in space and astrophysics
\cite{Alaoui2021apj,Yu-etal-FASS-2023,Roberg-Clark2019ApJ,
  Roberg-Clark2018prl}.

\section{supplemental material}
See Fig. \ref{Dxixi_forwardwhistlers} for the quasi-linear diffusion coefficient $D_{\xi\xi}$ at two different times showing the different harmonic
resonances connected to each other by the broad spectrum of
whistler waves. The excitation of secondary whistler waves by the backward diffusion further enhances the diffusion coefficient at the later time.
\setcounter{figure}{0}
\makeatletter 
\renewcommand{\thefigure}{S\@arabic\c@figure}
\makeatother

\begin{figure*}
\includegraphics[width=\textwidth]{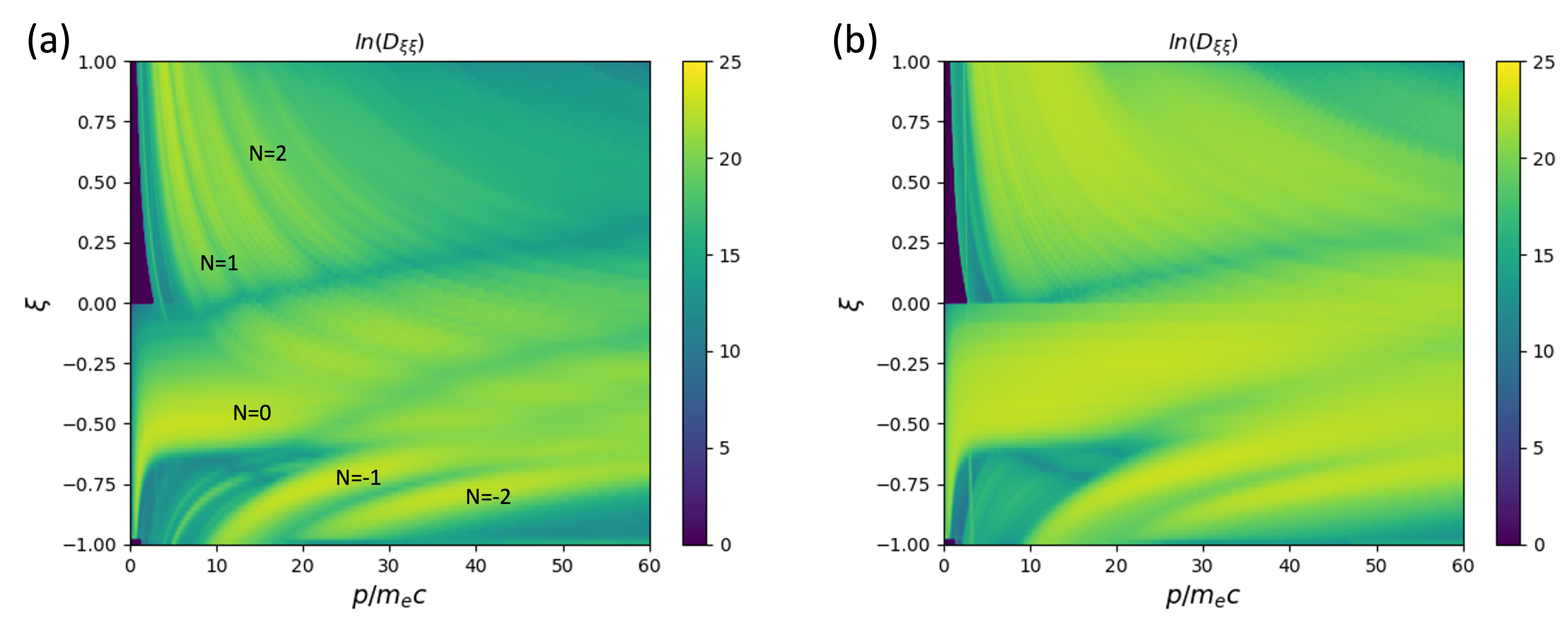}
\caption{Quasilinear diffusion coefficient $D_{\xi\xi}$ (arbitrary unit) in the momentum space calculated from the forward whistler branch ($kd_e\in [0,2]$) in the Fourier space, involving n=2 to -2 resonances for (a) $t\omega_{pe}=5000$ and (b) $t\omega_{pe}=250000$.}
\label{Dxixi_forwardwhistlers}
\end{figure*}

\textbf{Acknowledgment} We thank the U.S. Department of Energy Office
of Fusion Energy Sciences and Office of Advanced Scientific Computing
Research for support under the Tokamak Disruption Simulation and
SCREAM Scientific Discovery through Advanced Computing (SciDAC)
project, the Base Fusion Theory Program, and more recently the General
Plasma Science program, all at Los Alamos National Laboratory (LANL)
under contract No. 89233218CNA000001.  This research used resources of
the National Energy Research Scientific Computing Center, a DOE Office
of Science User Facility supported by the Office of Science of the
U.S. Department of Energy under Contract No. DE-AC02-05CH11231 using
NERSC award FES-ERCAP0028155 and the Los Alamos National Laboratory
Institutional Computing Program, which is supported by the
U.S. Department of Energy National Nuclear Security Administration
under Contract No. 89233218CNA000001.

%




\end{document}